\documentclass{ws-procs9x6}

\usepackage{psfrag,color}

\begin{document}

\title{Critical Casimir forces involving a chemically structured substrate}

\author{F.~Parisen~Toldin$^*$ and S.~Dietrich}

\address{
Max-Planck Institut f\"ur Metallforschung,\\
Heisenbergstr.~3,
D-70569, Stuttgart,
Germany, and\\
Institut f\"ur Theoretische und Angewandte Physik, Universit\"at Stuttgart,\\
Pfaffenwaldring 57,
D-70569 Stuttgart,
Germany\\
$^*$E-mail: parisen@mf.mpg.de\\
}

\begin{abstract}
Motivated by recent experiments with confined binary liquid mixtures near their continous demixing  phase transition we study the critical behavior of a system, which belongs to the Ising universality class, for the film geometry with one planar wall chemically structured such that there is a laterally alternating adsorption preference for the species of the binary liquid mixture.
By means of Monte Carlo simulations and finite-size scaling analysis we determine the critical Casimir force and the corresponding universal scaling function.
\end{abstract}


\bodymatter

\section{Introduction}
\label{sec:intro}
If a field exhibiting fluctuations with long wavelengths is confined between surfaces, long-ranged effective forces between them arise. The order parameter of a second-order phase transition represents such a field. The resulting effective force is known as the critical Casimir force. This phenomena, first predicted by Fisher and de~Gennes\cite{FG-78} is the analogue of the Casimir effect in quantum electrodynamics\cite{Casimir-48,Gambassi-09}. The critical Casimir force is characterized by a universal scaling function, which is determined by the bulk and surface universality classes (UC) of the confined system. It is independent of the microscopic details of the system, and it depends only on a few global and general properties, such as the number of components of the order parameter, the shape of the confinement, and the boundary conditions (b.c.) there.

In recent years, the critical Casimir effect has attracted numerous experimental and theoretical investigations; see Refs.~\refcite{Gambassi-09,Gambassi} for recent reviews. Wetting layers of binary liquid mixtures\cite{wetting} have provided indirect evidence and measurements of critical Casimir forces. Recently, a direct determination of the critical Casimir force has been reported\cite{HHGDB-08} using individual colloidal particles immersed in a binary liquid mixture close to its critical demixing point and exposed to a planar wall. 

Quantitatively reliable calculations of critical Casimir forces for laterally homogeneous b.c. have been obtained recently by means of Monte Carlo (MC) simulations. In this context, the Ising UC\cite{VGMD-07}, which describes the demixing transition in a binary liquid mixture, has been investigated and the critical Casimir force has been determined.

Experiments with binary liquid mixtures have also been used to study critical Casimir forces acting on a colloid in front of a chemically structured substrate\cite{SZHHB-08}, leading to a laterally varying adsorption preference. Such a system has been investigated theoretically for planar geometries within mean-field theory\cite{SSD-06} and for a curved geometry within the Derjaguin approximation\cite{TKGHD-09}.
Here we consider the Casimir force in the film geometry $L\times L_\parallel\times L_\parallel$, $L_\parallel\gg L$, such that the free energy density $\cal F$ per volume decomposes as
\begin{equation}
{\cal F}(\beta, L, L_\parallel) = f_\text{bulk}(\beta) + \frac{1}{L}f_\text{surf}(\beta) + \frac{1}{\beta}f_\text{ex}(\beta,L,L_\parallel),
\end{equation}
where $f_\text{bulk}(\beta)$ and $f_\text{surf}(\beta)$ are the bulk and the surface free energies per volume and area, respectively, in the thermodynamic limit $L_\parallel,L\rightarrow\infty$, and $f_\text{ex}$ is the excess free energy. The Casimir force $F_C$ per area and per $\beta^{-1}=k_BT$ is

\begin{equation}
F_C\equiv-\frac{\partial (Lf_\text{ex})}{\partial L}\Big|_{\beta,L_\parallel}.
\end{equation}

According to finite-size scaling\cite{Privman-89}, close to a continuous phase transition $F_C$ attains the following asymptotic scaling behavior:
\begin{equation}
\label{casimir}
F_C=\frac{1}{L^3}\theta\left(t(L/\xi_0^+)^{1/\nu},\rho\right),\qquad t\equiv (T-T_C)/T_C,
\end{equation}
where $T_C$ is the bulk critical temperature, $\rho\equiv L/L_\parallel$ is the aspect ratio, and $\xi_0^+$ is the non-universal amplitude of the correlation length $\xi=\xi_0^+t^{-\nu}$ in the disordered phase. The scaling function $\theta(x,\rho)$ is universal, i.e., it depends only on the UC of the bulk phase transition and on the surface UC, which is determined by the b.c. applied on the two surfaces. In the case of a confined binary liquid mixture, the order parameter $\phi$ is the difference between the local and the critical bulk concentration of one of the two components.
The substrates generically prefer one species over the other, so that $|\phi|$ is enhanced near the surface: this corresponds to the so-called extraordinary or normal surface UC\cite{Diehl-86} which, in a lattice model, can be implemented by external fields acting at the surfaces. For laterally homogeneous substrates, one has to distinguish between the case of surfaces with the same ($++$) and opposite ($+-$) adsorption preferences. For these systems laterally varying adsorption preferences can be realized\cite{SZHHB-08}.

\begin{figure}[t]
\begin{center}
\psfig{file=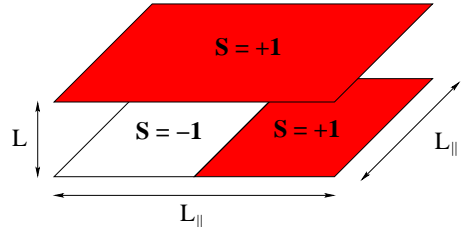,width=17em}
\caption{Film geometry with a single chemical step; S is the spin value at the surface.}
\end{center}
\vspace{-0.5em}
\label{lattice}
\end{figure}

Motivated by corresponding experimental results\cite{SZHHB-08}, we have performed MC simulations of a representative spin model on a lattice belonging to the Ising UC\cite{PTD-09}. As illustrated in Fig.~\ref{lattice}, the spins on the upper surface are fixed to $+1$. The lower surface is divided into two halves, one with spins fixed to $-1$ and the other with spins fixed to $+1$; the lattice constant is set to $1$. This mimics a single chemical step as the simplest lateral heterogeneity.
\section{Simulation method}
In order to calculate the Casimir force, we use the so-called coupling parameter approach\cite{VGMD-07}.
Given two systems with Hamiltonians ${\cal H}_0$ and ${\cal H}_1$ and the same configuration space, the free energy difference $\Delta F$ between the two systems can be computed as
\begin{equation}
\label{integral}
\Delta F = \int_0^1 d\lambda \langle \Delta {\cal H}\rangle_\lambda,\qquad \Delta {\cal H}\equiv {\cal H}_1-{\cal H}_0,
\end{equation}
where $\langle\ldots\rangle_\lambda$ denotes the thermal average over the ensemble described by the crossover Hamiltonian ${\cal H}_\lambda\equiv (1-\lambda){\cal H}_0+\lambda{\cal H}_1$, with $\lambda\in [0,1]$. This average can be evaluated by standard MC simulations, followed by numerically carrying out the integral in Eq.~(\ref{integral}). We have used Eq.~(\ref{integral}) by considering ${\cal H}_0$ as the Hamiltonian of the system shown in Fig.~\ref{lattice} and ${\cal H}_1$ as that of the corresponding film but with thickness $L-1$ and an additional, decoupled two-dimensional (2D) layer. One obtains
\begin{equation}
\label{I}
\frac{\Delta F}{L_\parallel^2} = -f_\text{bulk}(\beta) + f_\text{2D}(\beta) + \frac{1}{\beta}\frac{1}{L^3}\theta\left(t(L/\xi_0^+)^{1/\nu},\rho\right),
\end{equation}
where $f_\text{2D}(\beta)$ is the free energy density of the decoupled 2D layer and we have set the lattice constant to $1$, so that each term is dimensionless. We note that Eq.~(\ref{I}) is correct only up to corrections to scaling.

\section{Results}
At the bulk critical temperature the expression in Eq.~(\ref{I}) reduces to
\begin{equation}
\label{Icrit}
\frac{\Delta F}{L_\parallel^2} = -f_\text{bulk}(\beta_c) + f_\text{2D}(\beta_c) + \frac{1}{\beta_c}\frac{1}{L^3}\Theta\left(\rho\right),
\end{equation}
where $\Theta\left(\rho\right)\equiv\theta\left(0,\rho\right)$ is the critical Casimir amplitude. By fitting the MC data to this equation and by adding the expected additional scaling corrections, one can determine the critical Casimir amplitude\cite{PTD-09}.

\begin{figure}[t]
\begin{center}
\psfig{file=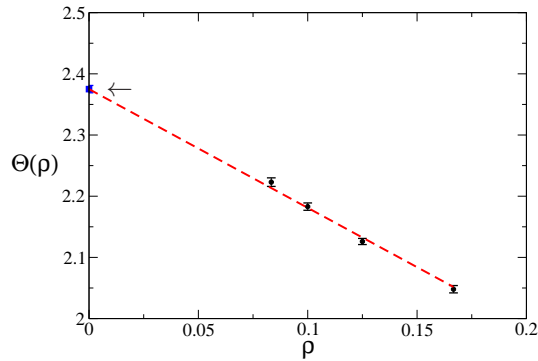,width=20em}
\caption{The amplitude $\Theta\left(\rho\right)$ of the critical Casimir force at $T_c$ for the system shown in Fig.~\protect\ref{lattice} as a function of the aspect ratio $\rho=L/L_\parallel$.}
\end{center}
\label{critvsrho}
\end{figure}

\begin{figure}
\begin{center}
\psfig{file=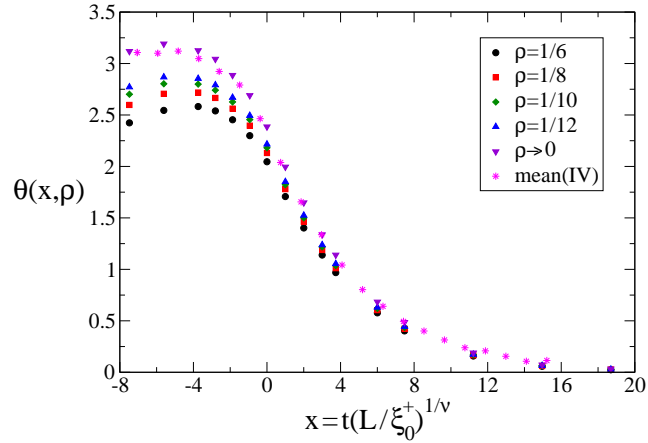,width=24em}
\caption{The universal scaling function of the critical Casimir force for the system shown in Fig.~\protect\ref{lattice} and for various aspect ratios $\rho=L/L_\parallel$. We also show the scaling function obtained as the mean value of those for laterally homogeneous ($++$) and ($+-$) b.c. based on the approximant IV presented in Ref.~\protect\refcite{VGMD-07}. Error bars are smaller than the symbol size.}
\end{center}
\vspace{-0.7em}
\label{theta}
\end{figure}

As shown in Fig.~\ref{critvsrho} the Casimir amplitude varies linearly for small $\rho$.
We find $\Theta(1/6)=2.048(6)$, $\Theta(1/8)=2.126(5)$, $\Theta(1/10)=2.183(6)$, $\Theta(1/12)=2.223(7)$, and $\Theta(\rho\rightarrow 0)=2.386(4)$.
In the limit $\rho\rightarrow 0$ the force is expected to be the average of the force for laterally homogeneous ($++$) and ($+-$) b.c. At criticality\cite{VGMD-07} $(\Theta_{++}+\Theta_{+-})/2=2.33(4)$, in agreement with our results. The calculation of the full scaling function $\theta(x,\rho)$ requires the subtraction of the $L-$independent terms in Eq.~(\ref{I}) (see Ref.~\refcite{PTD-09} for details). In Fig.~\ref{theta} we report the resulting scaling function for various aspect ratios. We also show a comparison with the scaling function obtained as the average of those for homogeneous ($++$) and ($+-$) b.c., based on the data of Ref.~\refcite{VGMD-07}. For $\rho\rightarrow 0$ we find good agreement with our results.

\psfrag{QQQ}[2]{$\leftarrow$}

\end{document}